\documentclass[twocolumn,prl,showpacs,amsfonts,superscriptaddress,floatfix,nofootinbib,10pt]{revtex4-1}
\usepackage{amsmath}
\usepackage{graphicx}
\usepackage{footmisc}
\pdfoutput=1
\usepackage{color}

\newcommand{\ii}{\mathrm{i}}

\newcommand{\vect}[1]{{\mathbf{#1}}}
\newcommand{\fref}[1]{figure~\ref{#1}}

\begin{document}

\title {Chern insulator with large Chern numbers. Chiral Majorana fermion liquid}
\author{Igor N. Karnaukhov}
\affiliation{G.V. Kurdyumov Institute for Metal Physics, 36 Vernadsky Boulevard, 03142 Kiev, Ukraine}
\begin{abstract}
In the framework of Hofstadter`s approach we provide a detailed analysis of a realization of exotic topological states such as the Chern insulator (CI) with large Chern numbers. In a transverse homogeneous magnetic field a one-particle spectrum of fermions transforms to an intricate spectrum with a fine topological structure of the subbands.  In a weak magnetic field $H$ for a rational magnetic flux, a topological phase with  a large Chern number is realized near the half filling. There is an abnormal behavior of the Hall conductance ${\sigma_{xy}\simeq \frac{e}{2 H}}$. At half-filling, the number of chiral Majorana fermion edge states increases sharply forming a new state, called the  chiral Majorana fermion liquid.
\end{abstract}
\pacs{}
\maketitle

\section{Introduction}

In the case of a rational magnetic flux through the unit cell ${\phi=p/q}$ ($p$ and $q$ are coprime integers), a new physics of the 2D model systems subjected to a perpendicular magnetic field arises from its two magnetic and lattice competing scales~\cite{Hof}. A~strong periodic potential leads to an intricate spectrum of the topological system, the problem brings to play the commensurability of these two scales~\cite{Har}. For the particular case, when the magnetic flux  assumes either the continued fraction approximations towards the golden mean or the golden mean itself, the multifractal properties of Harper`s equation, the winding numbers were discussed in  \cite{5}.
The problem of realization of the Hofstadter Hamiltonian with ultracold atoms in optical lattices is addressed, for example, in Refs. \cite{a2,a3}

Wiegmann and Zabrodin pointed out \cite{2} that Hofstadter`s model responsible is closely related to the quantum group $U_q(sl_2)$,  the model Hamiltonian is determined in terms of the generators of the quantum group (see also the approach of Faddeev and Kashaev \cite{3}). Using the exact solution of the Hofstadter model \cite{2,3}
for the semi-classical limit at $p=1$ and  $q \to \infty$, the behavior of the wavefunction was calculated at zero energy (for the center of the spectrum) $|\psi_j |^2= G_j=\frac{2}{\sin(\pi (j-1/2)/q)}$ ($j$ is the site of the lattice) \cite{4}. Near the edge $j=0$, the finite size correction is given by $|\psi_{2j+1} |^2 = Q(j)G_{2j+1}$, with $ Q(0) =\frac{\pi}{4}$, $ Q(1)=\frac{5\pi}{16}$, $Q(2) =\frac{81\pi}{256}$. The authors of \cite{4} noted that power-low behavior of $|\psi_j |^2$ is critical and unnormalizable. For the golden mean flux, the wavefunction is multifractal and critical, has a clear self-similar branching structure. Harper \emph{et~al.}~\cite{H} shown that the Hofstadter model on a square lattice converges to continuum Landau levels in the limit of small flux per plaquette \mbox{(in the ${q \to \infty}$ limit)}.

Filled $r$ bands with Chern numbers~$C_\gamma$ ($1\leq \gamma\leq r$) yield a Hall conductance
$\sigma_{xy}= \frac{e^2 }{h}|\sigma_r|, \sigma_r=\sum_{\gamma=1}^r C_\gamma $. The relationship between Hall conductance $\sigma_{xy} $ and energy spectrum, the Chern numbers of isolated subbands are discussed in \cite{6}.  Dana \emph{ et al.} \cite{D}  proposed  that magnetic translational symmetry yields the diophantine equation $p |o_r| + 2 q s = 2 r$. The solution of the equation can be uniquely determined by imposing the condition $|\sigma| \leq q$, that is realized in the model.

The dynamics of a charged particle in a magnetic field applied perpendicular to the plane of the lattice and to the electric field in the plane of the lattice is considered in \cite{a1}. A non-zero magnetic field crucially modifies this dynamics of fermions along the edges.
Exact results \cite{2,3,7} and numerical calculations \cite{9,8} show that the behavior 2D fermion systems in a transverse magnetic field does not depend on the symmetry of the lattice.

Nontrivial topological order in a band is indicated by non-zero Chern number.
The~topological insulating phase state is determined by the Chern number of all fermion bands below the Fermi energy.
The~Chern number is a topological invariant which can be easily defined for the $\gamma$-band isolated from all other bands by the formula
\begin{equation}
    C_\gamma =\frac{1}{2\pi}\int_{BZ}{\cal B}_\gamma (\vect{k}) \mathrm{d}^2\vect{k}
    \label{eq:Chern}
\end{equation}
integrating the Berry curvature ${\cal B}_\gamma (\vect{k}) = \nabla_{\vect{k}}\times {\cal A}_\gamma (\vect{k})$ over the Brillouin zone~(BZ). The Berry potential
$${\cal A}_\gamma (\textbf{k})= -\ii \langle u^\gamma(\vect{k}) | \nabla_\vect{k} | u^\gamma(\vect{k}) \rangle $$
is defined in terms of the Bloch states $u^\gamma (\textbf{k})$.
If~bands in the set are isolated, $\displaystyle C_{\Gamma} = \sum_{\gamma\in\Gamma} C_\gamma$.

\section{The model}

We will analyze a model of the 2D CI determined in the framework of Hofstadter`s approach \cite{Hof}.
In the presence of a transverse homogeneous magnetic field $H\vect{e}_z$ the model Hamiltonian is determined according to \cite{Hof}
\begin{multline}
    {\cal H} =  \sum_{n,m} [t^x(n,m) a^\dagger_{n,m}  a_{n+1,m} +\\ t^y(n,m)a^\dagger_{n,m}  a_{n,m+1} + Hc]
    \label{eq:H}
\end{multline}
where $a^\dagger_{n,m} $ and $a_{n,m}$ are the spinless fermion operators on a site $\{n,m\}$  with the usual anticommutation relations. The Hamiltonian describes the next-nearest hoppings along the $x$-direction with the magnitude $t$ ($t^x(n,m)=t$) and along the $y$-direction with the hopping integral $t^y (n,m)=\exp[2i \pi (n-1)\phi]$,
where $\phi = \Phi / \Phi_0$ is a magnetic flux ${\Phi=H}$ through the unit cell measured in quantum flux unit ${\Phi_0=h/e}$, a homogeneous field~$H$ is represented by its vector potential ${\vect{A} = H x \vect{e}_y}$. The case t=1 corresponds  the original Hofstadter model \cite{Hof}.
We focus on the 2D system in the form of a hollow cylinder, with periodic boundary conditions along the $ y $-direction and size $ N $ along the $x$-direction.

\section{Topological structure of the spectrum }
\subsection{The flux $\phi =\frac{1}{3}$}

\begin{figure}[tp]
    \centering{\leavevmode}
     \centering{\leavevmode}
    \begin{minipage}[h]{,475\linewidth}
\center{\includegraphics[width=\linewidth]{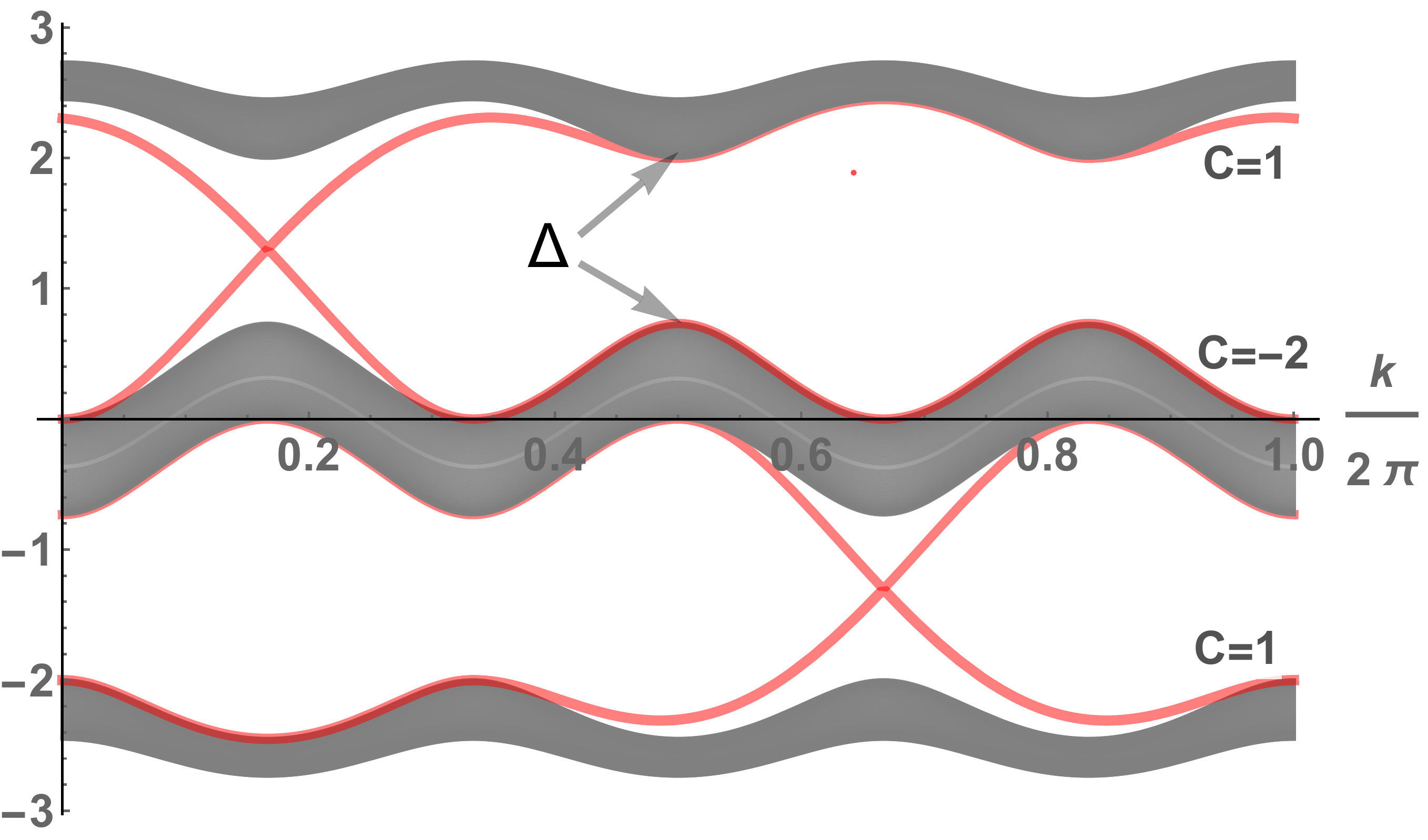}\\ a) }
\end{minipage}
\begin{minipage}[h]{.475\linewidth}
\center{\includegraphics[width=\linewidth]{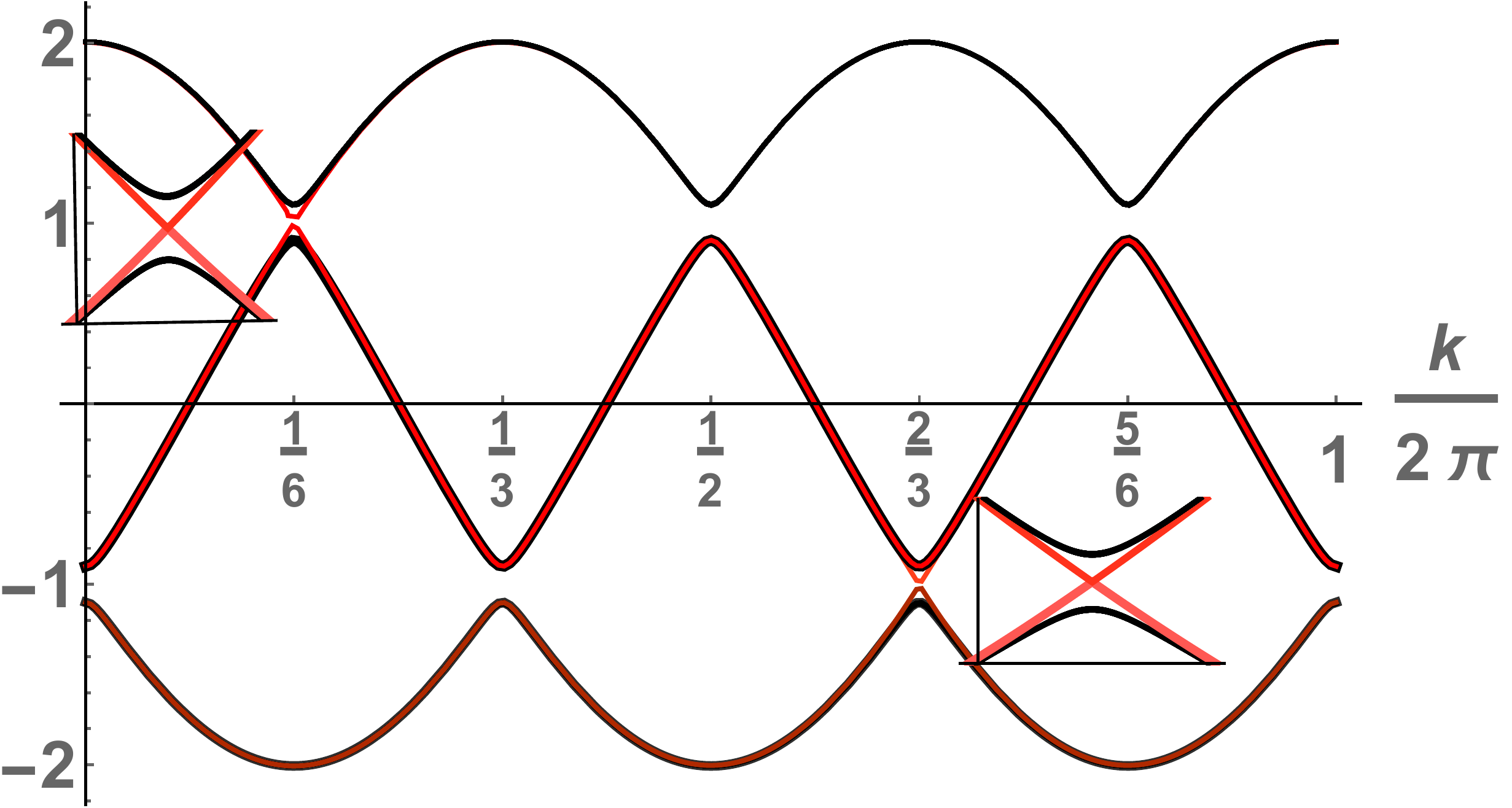}\\ b) }
\end{minipage}
\begin{minipage}[h]{.475\linewidth}
\center{\includegraphics[width=\linewidth]{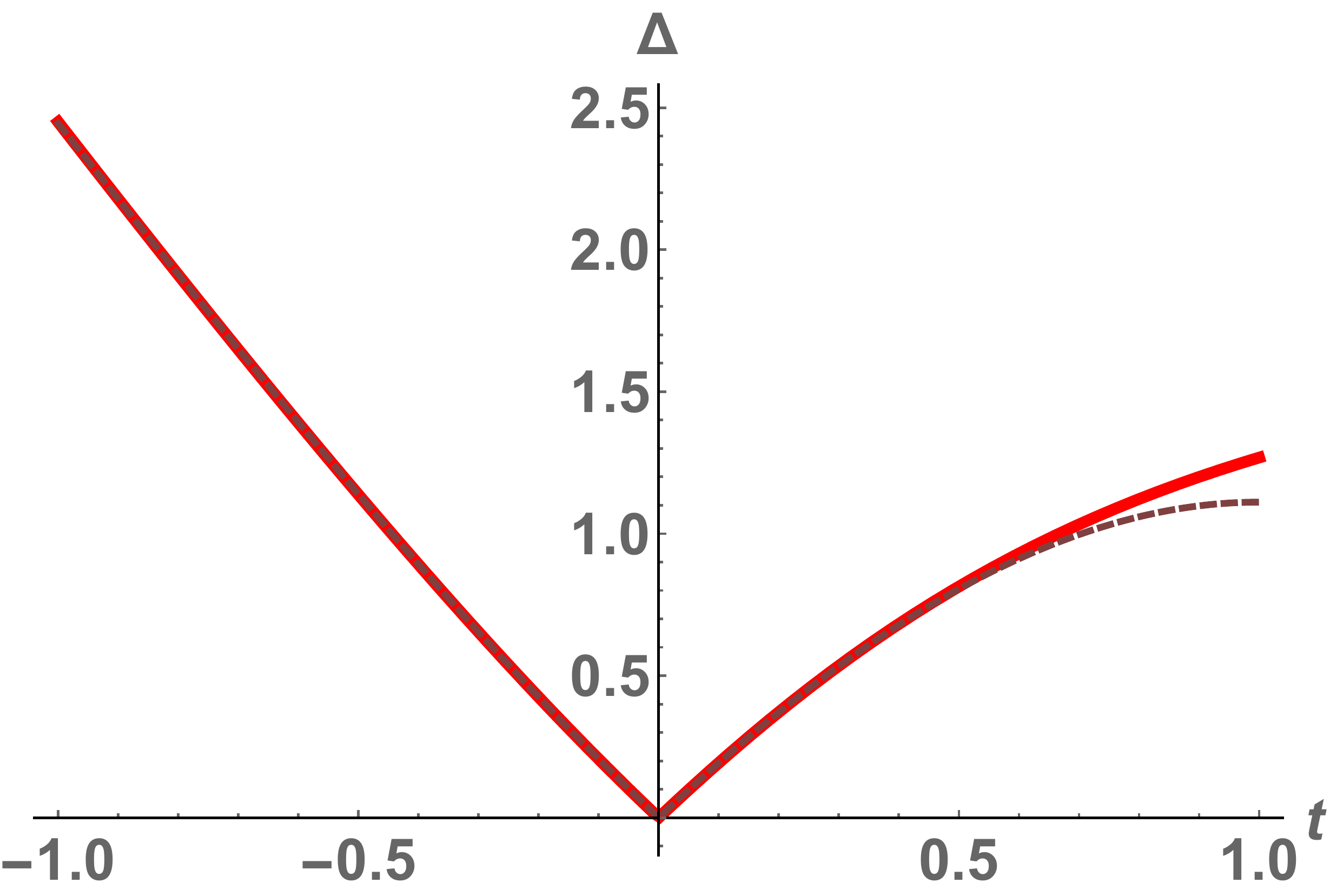}\\ c)}
\end{minipage}
\begin{minipage}[h]{.475\linewidth}
\center{\includegraphics[width=\linewidth]{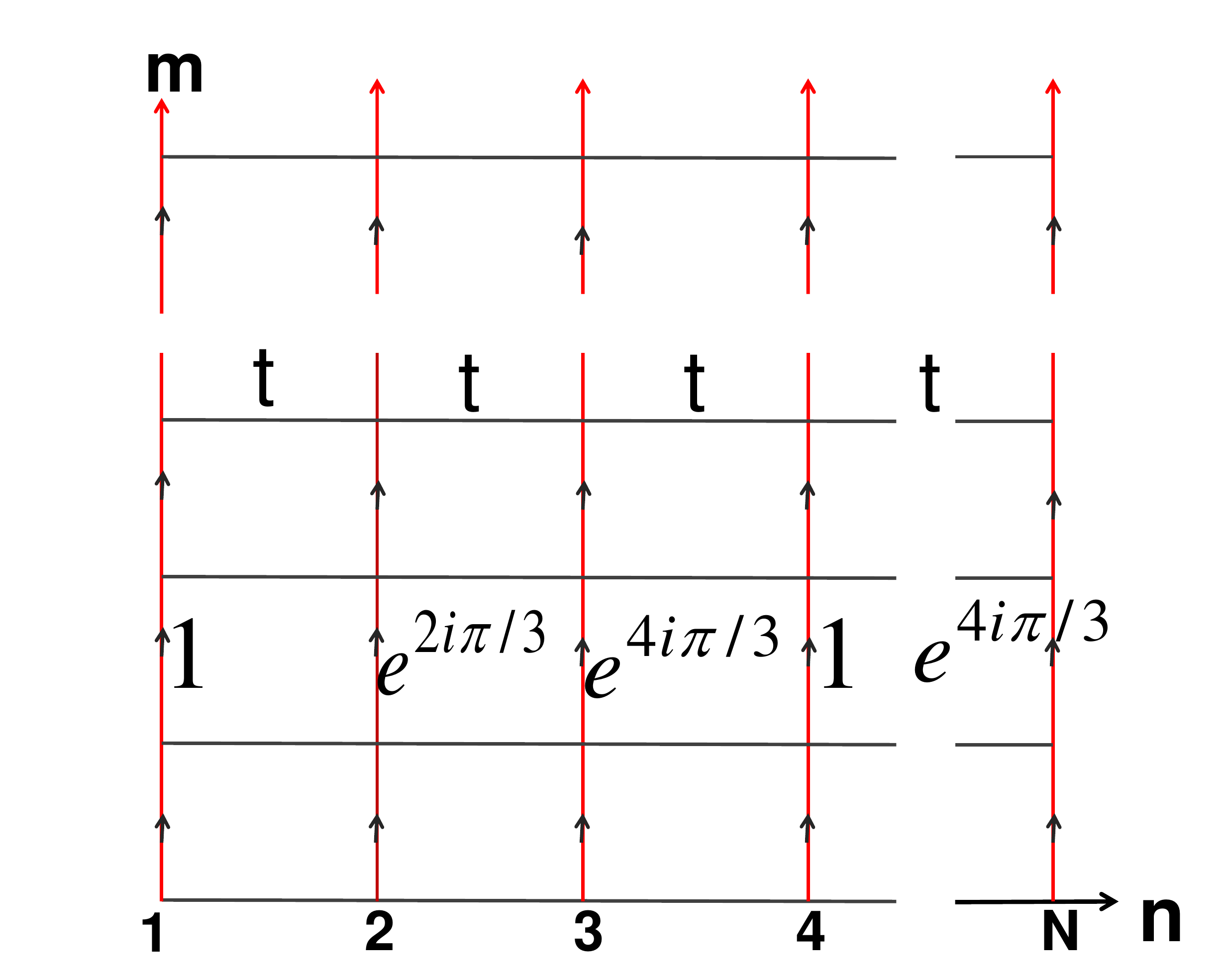}\\ d)}
\end{minipage}
\caption{(Color online)
Energy levels  calculated on a cylinder with open boundary conditions along $y$-direction for $t=1$ a), $t=\frac{1}{10}$ b) at $\phi=\frac{1}{3}$, the insets zoom the regions with the edge modes around $\{\frac{\pi}{3},1\}$ and $\{\frac{4\pi}{3},-1\}$. The value of gaps that separated the band on the isolated topological subbands, as a functions of $t$ (red line) and its asymptotic at small $t$ $\Delta \simeq|2 t|(1 - \frac{t (3 + t)}{9})$ (dotted line) c). The hopping integrals are determined for $q=3$, the fermion chains are indicated in red color d).
  }
\label{fig:1}
\end{figure}

\begin{figure}[tp]
    \centering{\leavevmode}
     \centering{\leavevmode}
    \begin{minipage}[h]{.9\linewidth}
\center{\includegraphics[width=\linewidth]{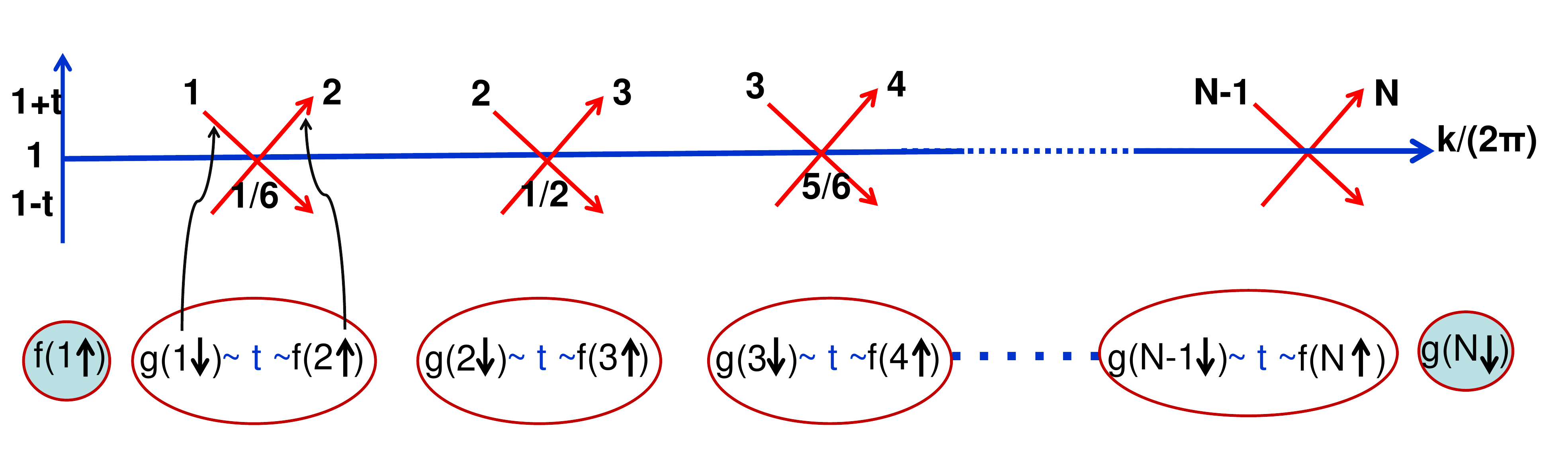} 
}
\end{minipage}
\caption{(Color online) Pairing of Majorana fermions in the Hamiltonian (\ref{eq:H1}) and forming chiral edge states $f(1,k$) and $g(N,k)$. 
}
\label{fig:2}
\end{figure}

We consider an evolution of the one-particle spectrum of the Hamiltonian \eqref{eq:H} in a transverse homogeneous magnetic field with rational magnetic flux $\phi= \frac{p}{q}$. In the case $q>2$ the magnetic field breaks a time reversal symmetry \cite{K1,K2}, leads to topological states of fermions \cite{K3}. Our~starting point is $q=3$, when the one-particle spectrum consists of three topologically nontrivial subbands with the Chern numbers $\{1,-2,1\}$  (see in \fref{fig:1} a),b)). A detailed calculation of the Chern number for a rational flux is given in \cite{d,d1}. In an external field the spectrum of fermions is intricate, the band is split into the topological subbands $\gamma$ with non-trivial topological index $C_\gamma$. The structure and number of subbands in the fine structure of the spectrum depend on the value of the magnetic flux. The Chern number corresponds to the total number of edges modes with respect to their chirality, even for groups of intersecting bands.
The excitation spectrum of the sample in stripe geometry consists of the chiral edge modes localized at the boundaries, which are indicated in red color (see in \fref{fig:1} a),b)). The structure of the spectrum does not depend on the value of $t$, the gaps close at $t=0$ and open for $t\neq 0$.  The values of gaps are equal to $\Delta = \frac{3}{2} |1+t-\sqrt{1+t^2-2t/3}| $. The value of $\Delta$ is shown in \fref{fig:1}c). The topology of the spectrum is not changed at the point $t=0$. The gaps close at $t=0$, but the spectrum of the excitations has not the Dirac-type in the $k_x$-direction (\textbf{k} is a wave vector).

Let us consider the spectrum of the excitations in the case of a weak coupling along the \emph{x}-direction at small values of $t$ (see in \fref{fig:1}b)). In the $t \to 0$ limit the fermions form the noninteracting chains along the \emph{y}-direction (see in \fref{fig:1}d)). The energies of fermion excitations in the chains intersect at the energies equal to $\pm 1$ for $k_y= \frac{p \pi}{3}$ $p=1,...,6$ (see in \fref{fig:1}b). Due to the hoppings of fermions between the nearest chains $t$ the gaps open at the energies equal to $\pm 1$. The tunneling of fermions between chains dominates in the regions of the crossings of energies of the nearest chains, since the conservation of the energy and the momentum of fermions are realized automatically. The Hamiltonian, which takes into account the low energy excitations of Majorana fermions, is determined as follows
\begin{equation}
{\cal H}^{I}_{eff} = i t  \sum_{n=1}^{N-1} g(n,k_{n,n+1})f(n+1,k_{n,n+1})\\
    \label{eq:H1}
\end{equation}
where $k_{1,2}=\frac{2\pi}{3}$ or $\frac{5\pi}{3}$, $k_{2,3}=0$ or $ \pi$, $k_{3,1}=\frac{4\pi}{3}$ or $\frac{\pi}{3}$, $1,2,3$ numerate the chains for $q=3$ (the sets values of $k$ for the energies $-1$ or $1$, see in \fref{fig:1} b).  Majorana operators $g(n,k)$, $f(n,k)$ describe the chiral modes (with opposite velocities) on the chain $n$, Majorana state operators $\gamma (j)$  defined by the algebra $\{\gamma (j),\gamma (i)\}=2 \delta_{j,i}$ and $\gamma (j)=\gamma^\dagger (j)$.

Chiral Majorana fermions from different nearest chains are paired together in the Hamiltonian \eqref{eq:H1}, Majorana fermions form noninteracting dimers along the $x-$direction as noted in \fref{fig:2}. According to Kitaev \cite{Ki} the Hamiltonian (\ref{eq:H1}) is determined as ${\cal H}^I_{eff} = t  \sum_{n=1}^{N-1} (2 c^\dagger_{n,k} c_{n,k}-1)$, where $c^\dagger_{n,k}, c_{n,k}$ are the Fermi operators constructed from the two Majorana bound states located at the nearest $y-$chains chains (see in \fref{fig:2}). The Kitaev chain Hamiltonian  \cite{Ki} describes topological superconductivity in a chain of spinless fermions.

The operators $f(1,\frac{\pi}{3})$ and $g(N,\frac{\pi}{3})$ are free Majorana fermions at the energy equal to $1$, they remain unpaired and form edge states.
The gapless edge modes existing in the energy range of about $ -1 $ are associated with the Majorana operators $g(1,\frac{4\pi}{3})$ and $f(N,\frac{4\pi}{3})$ (see in \fref{fig:1} b)).
They are localized near the boundaries of the sample \cite{Ki}. The chiral gapless edge modes do exist in the gaps, connect the lower and upper fermion subbands (see in \fref{fig:1} a),b)).

\subsection{The flux $\phi=\frac{1}{5}$, rational flux }

\begin{figure}[tp]
    \centering{\leavevmode}
     \centering{\leavevmode}
   \begin{minipage}[h]{.475\linewidth}
\center{\includegraphics[width=\linewidth]{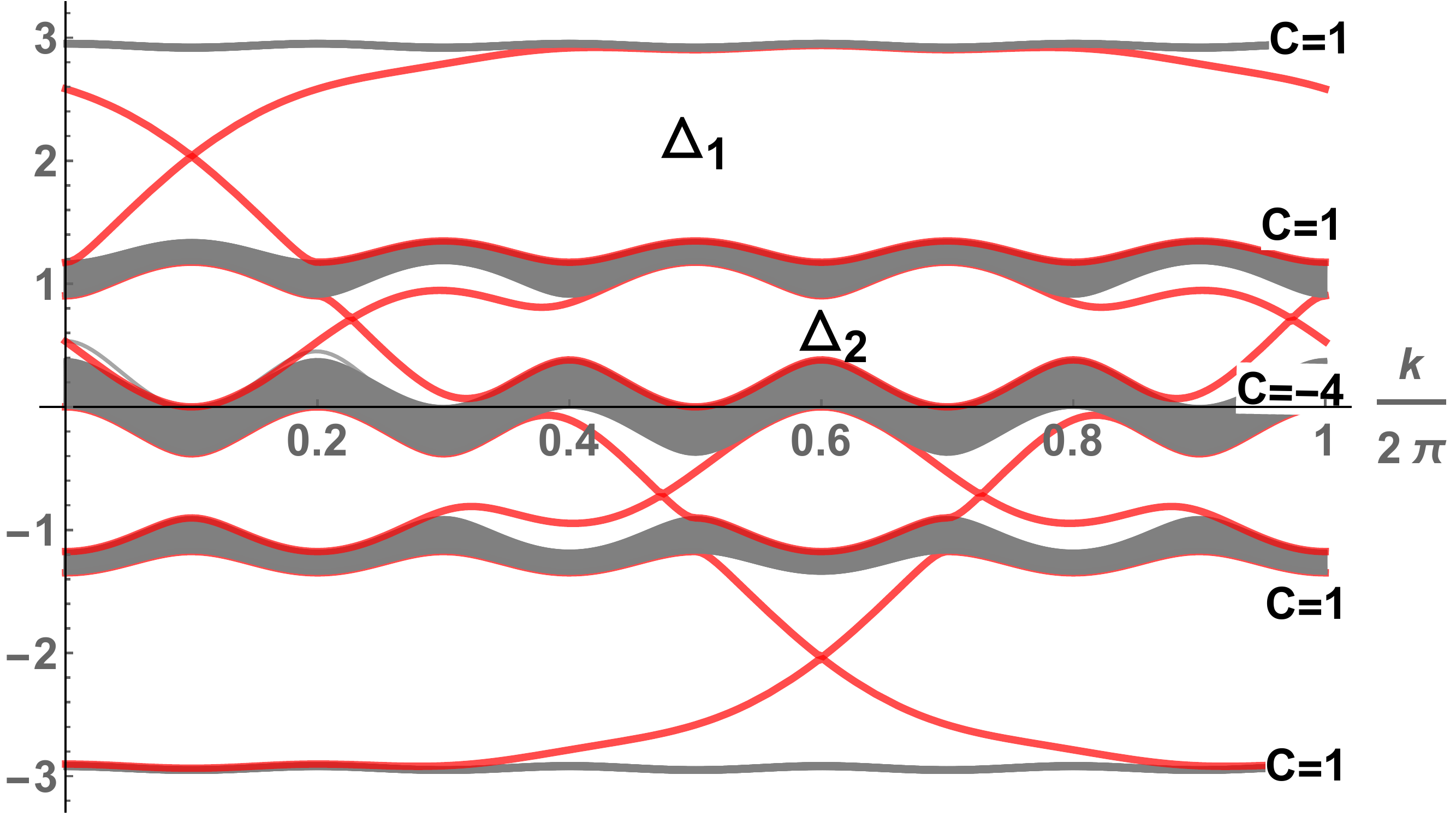}\\ a)}
\end{minipage}
\begin{minipage}[h]{.475\linewidth}
\center{\includegraphics[width=\linewidth]{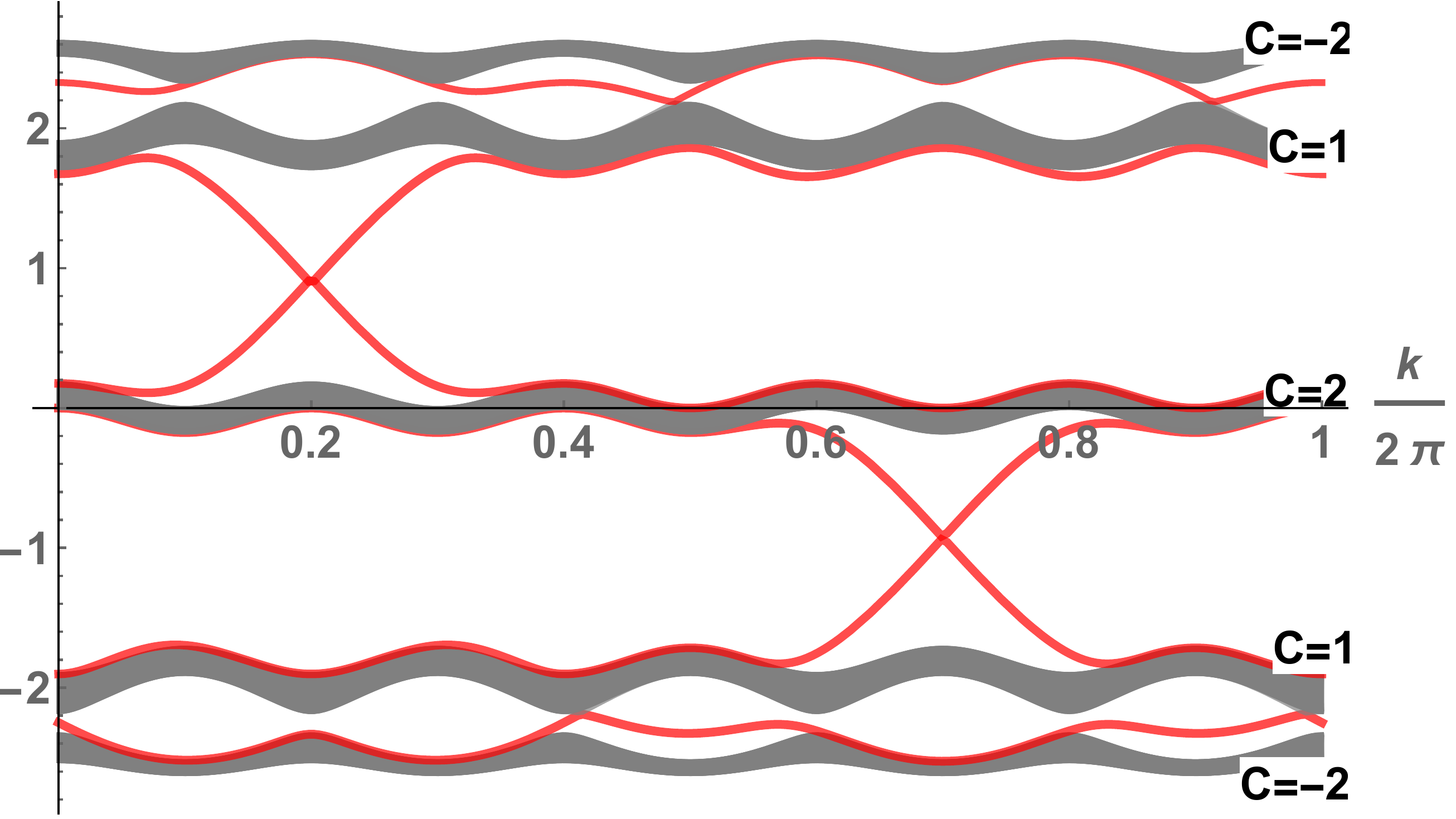}\\ b)}
\end{minipage}
\begin{minipage}[h]{.6\linewidth}
\center{\includegraphics[width=\linewidth]{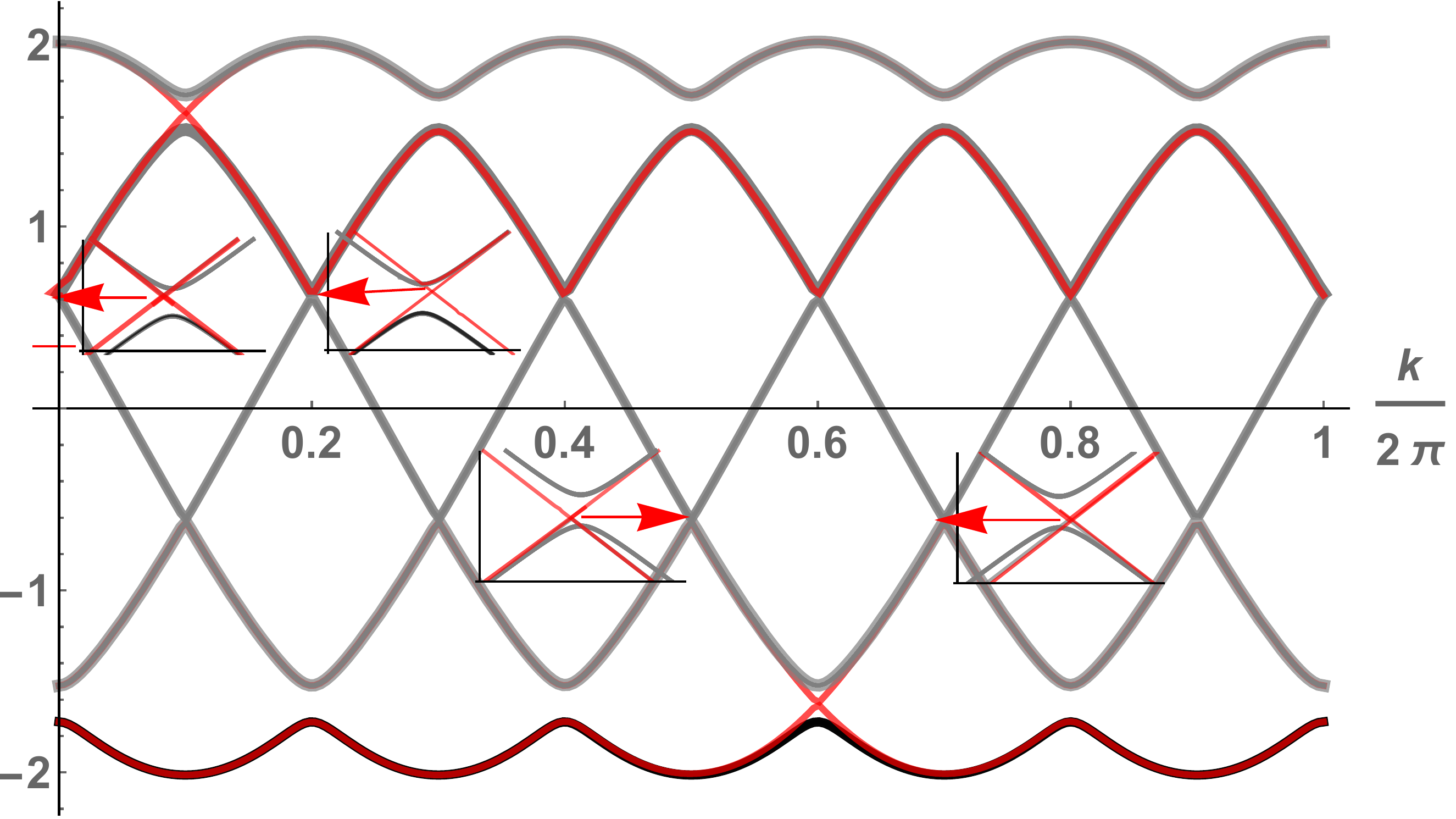}\\ c) }
\end{minipage}
\begin{minipage}[h]{.35\linewidth}
\center{\includegraphics[width=\linewidth]{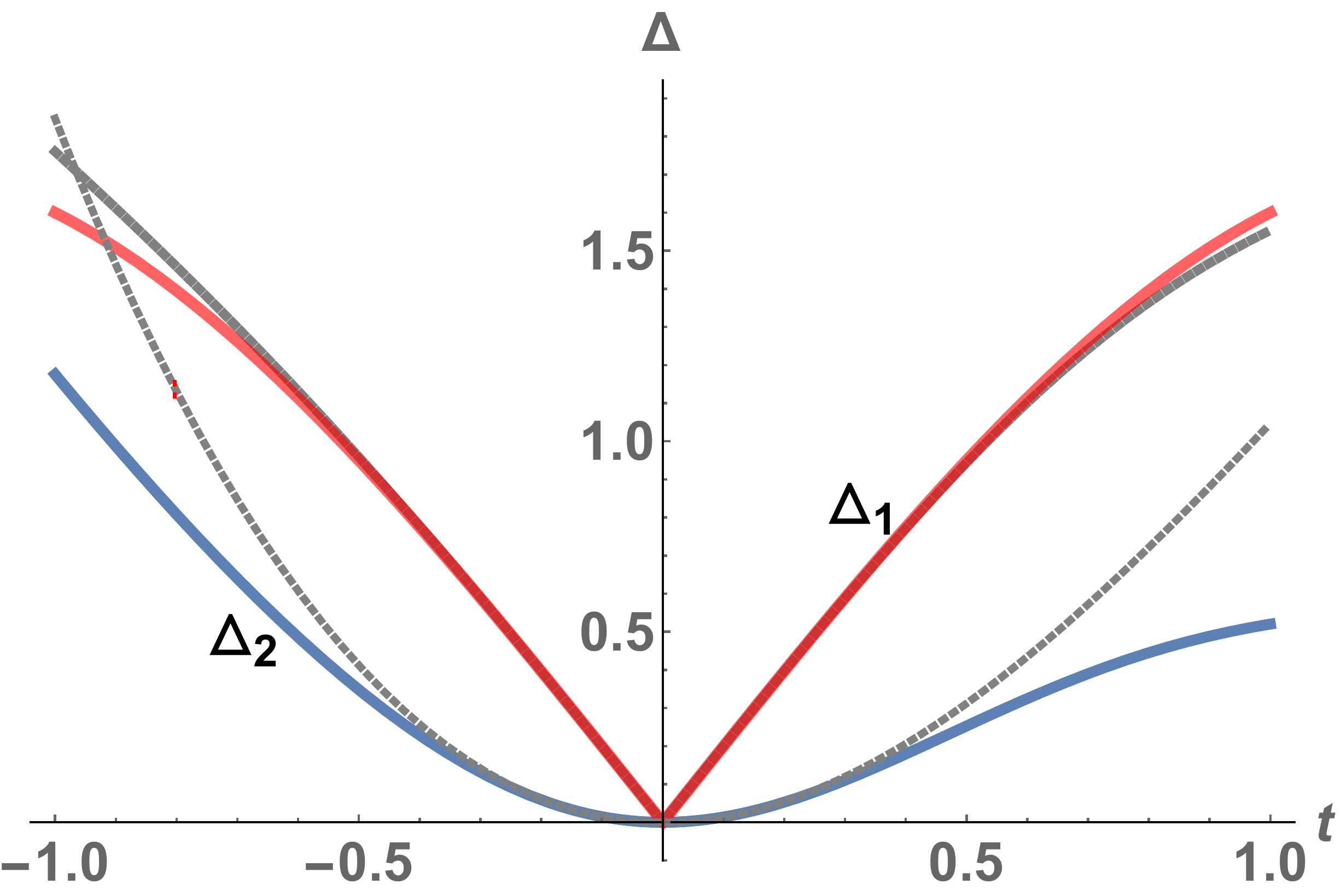}\\ d)}
\end{minipage}
\caption{(Color online)
Energy levels  calculated as in Fig.1 for $\phi=\frac{1}{5}$ at $t=1$, $p=1$ a), $t=1$, $p=2$ b) and $t=\frac{1}{10}$, $p=1$ c). The values of gaps, separated the bands on the isolated topological subbands, as the functions of $t$ d) ($\Delta_1$ red, $\Delta_2$ blue lines) with the asymptotics at a weak coupling (dotted lines) $\Delta_1\simeq |2t |(1-\frac{t^2}{5})$ and $\Delta_2\simeq (1+\frac{1}{\sqrt{5}})t^2 - \frac{2 t^3}{5}$.
  }
\label{fig:3}
\end{figure}

The spectrum of the excitations contains five sobbands and four gaps, which are determined by two values of-$\Delta_1$ and $\Delta_2$ (see in Fig.\ref{fig:3}a)), with different structure of edge modes in the regions of the gaps. At the energies equal to $\pm 2 \cos \frac{\pi}{5}$  the edge states with two chiral modes $f(1,\frac{\pi}{5})$ and $g(N,\frac{\pi}{5})$ into the gap $\Delta_1$ are described by the Hamiltonian (\ref{eq:H1}). The gap $\Delta_2$ is defined by an effective hopping of Majorana fermions, located at the next-nearest chains along the $y$-direction, with the following effective low energy Hamiltonian
\begin{equation}
{\cal H}^{II}_{eff} = i \tau  \sum_{n=1}^{N-2} g(n,k_{n,n+2})f(n+2,k_{n.n+2}),
 \label{eq:H2}
\end{equation}
where in a weak coupling  $\tau \simeq t^2$ or according to numerical calculations $\tau =(1+\frac{1}{\sqrt{5}})t^2$,
$k_{1,3}=\frac{3\pi}{5}$ or $\frac{8\pi}{5}$,
$k_{2,4}=\frac{\pi}{5}$ or $\frac{6 \pi}{5}$,
$k_{3,5}=\frac{9\pi}{5}$ or $\frac{4\pi}{5}$,
$k_{4,1}=\frac{7\pi}{5}$ or $\frac{2\pi}{5}$,
$k_{5,2}=\pi$ or $0$ for the energies $\mp 2 \cos \frac{2\pi}{5}$.
The Hamiltonian \eqref{eq:H2} describes the dimers that contain Majorana fermions located at the next-nearest neighbor sites in the $x-$chains. Majorana fermions $f(1,\frac{2\pi}{5})$ and $g(2,\frac{2\pi}{5})$ and $g(N,0)$  and $f(N-1,0)$ are free Majorana fermions, they remain unpaired and form edge states into the gap $\Delta_2$ at the energy $2 \cos \frac{2\pi}{5}$ (see in \fref{fig:3}c)).

When the flux is rational, the Hamiltonian \eqref{eq:H} admits a periodic representation on a magnetic unit cell comprising $q$ cells along the $x$-direction. We~use Bloch's theorem and calculate the Chern numbers of the isolated subbands.
The topological picture of model's spectrum is defined by the Chern numbers of each subband $C_{\gamma}$.
The Chern numbers $\{1,1,-4,1,1,\}$ for $p=1$ and $\{-2,1,2,1,-2\}$ for $p=2$ determine the topological structure of the spectrum at $q=5$.
The following relations follow from symmetry and numerical calculations:
$\sum_\gamma C_{\gamma}=0$ and $C_{\gamma}(\phi)=-C_{\gamma}(-\phi)$.
The Chern numbers are defined for isolated subband for a rational flux $\phi= \frac{p}{q}$, $C_{\gamma}( \frac{p}{q})=-C_{\gamma} (\frac{q-p}{q})$,
$p$ and $q$ are integers, $q$ enlarges a magnetic unit cell and thus defines the number of subbands $\gamma =\overline{1,\ldots,q}$. From numerical calculations and exact solution \cite{2}, it follows that for $p=1$ and odd $q=2s+1$ all
subbands are isolated by the gaps, therefore  $C_{\gamma}$ is defined for arbitrary~$\gamma$.
Useful relations are obtained for sets values of rational fluxes:
${|C_{s+1}(\tfrac{1}{2s+1})| = 2s}$, ${|C_{s+1}(\tfrac{2}{4s\pm1})| = 2s}$ and $|C_{s+1}(\tfrac{s}{2s+1})| = 2$.
The~Chern number of the middle subband for given $q=2s+1$  is varied from the maximal value $q-1$ at $p=1$ to the minimal value 2 at $p=(q-1)/2$, $2\leq |C_{s+1}| \leq q-1$.
The~phase state of CI with a large Chern number and corresponding Hall conductivity $\sigma_{xy}=(q-1)\frac{e^2}{2 h} \simeq \frac{e}{2 H}$
is realized in a weak magnetic field in a narrow region near the half filling when filling $\frac{1}{2} (1\pm \frac{1}{q})$.
Gapless chiral edge modes are localized in a wide region near the boundaries, from 1 (N) to $\frac{q-1}{2}$ ($N-\frac{q-3}{2}$) lattice cites. There are two different separated regions of the phase states of fermions:
itinerant fermions in the bulk at $\frac{q-1}{2}<n< N-\frac{q-3}{2}$ and localized chiral gapless edge Majorana state at $ 1\leq  n\leq \frac{q-1}{2}$ and $N-\frac{q-3}{2}\leq n \leq N$. In the limit of a weak magnetic field, the region of the existence of the localized chiral gapless edge Majorana state  has a large size $\sim \frac{1}{H}$.

\begin{figure}[tp]
    \centering{\leavevmode}
     \centering{\leavevmode}
   \begin{minipage}[h]{.6\linewidth}
\center{\includegraphics[width=\linewidth]{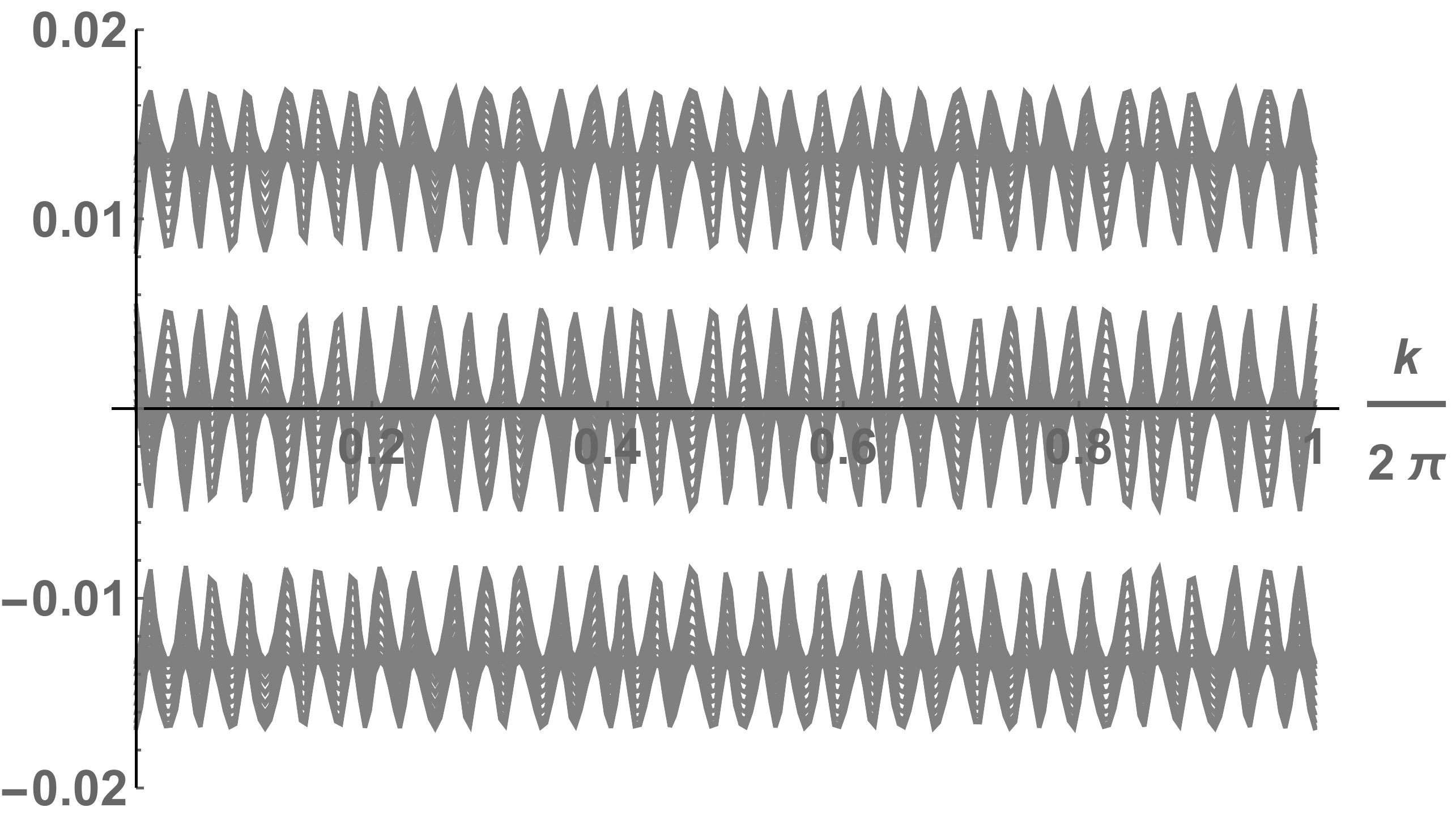}\\ a)}
\end{minipage}
\begin{minipage}[h]{.6\linewidth}
\center{\includegraphics[width=\linewidth]{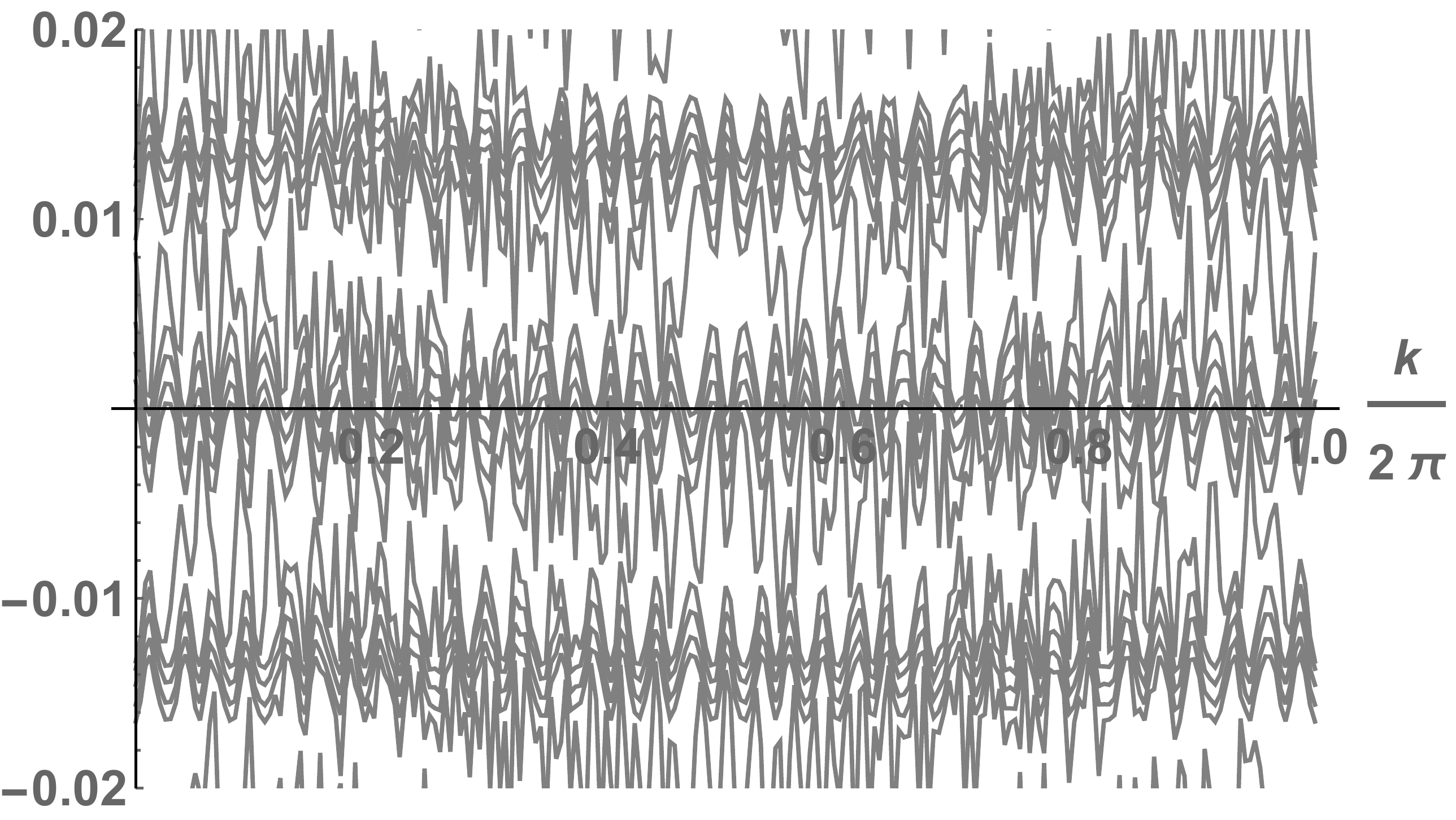}\\ b)}
\end{minipage}
\begin{minipage}[h]{.6\linewidth}
\center{\includegraphics[width=\linewidth]{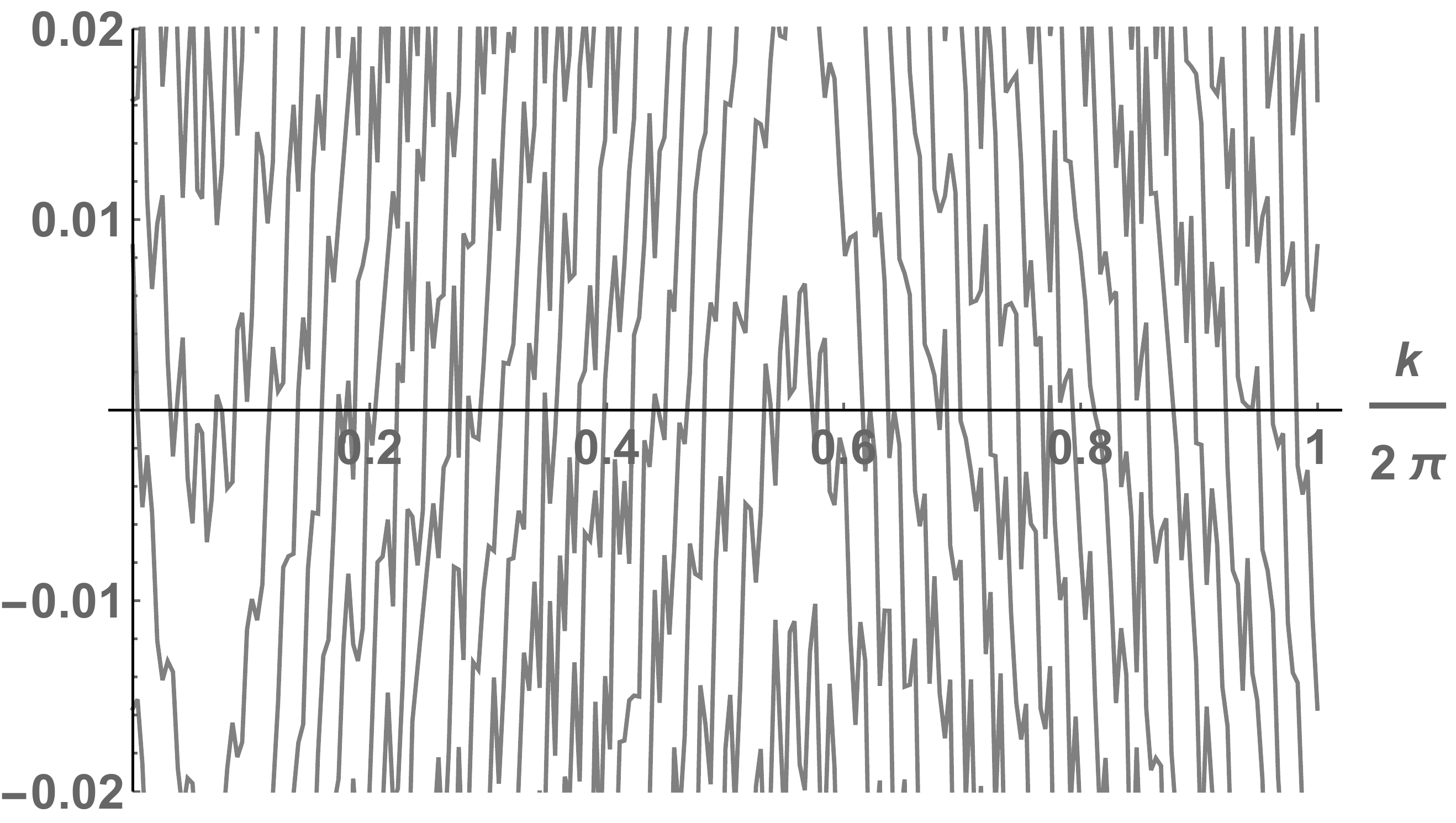}\\ c)}
\end{minipage}
\caption{
A low energy spectrum (a hyperfine structure of middle subband), calculated for flux $\phi=\frac{1}{201}$ at $t=1$:
 with periodic boundary conditions a), with open boundary conditions along the $y$-direction at $N=1005$ b) and at $N=177$ c), $k$ is the wave vector along the $y$-direction.
  }
\label{fig:4}
\end{figure}

Figure \ref{fig:4}, which shows the calculations of a hyperfine structure of the middle subband, illustrates what was said above. In \fref{fig:4}a) the center of the spectrum is calculated using periodic boundary conditions for flux $\phi=\frac{1}{201}$ at $t=1$, when the fermion states are determined by Bloch`s function. Well-defined subbands separated by the gaps, here the wave vector directed along the $y$-direction and $k_x$ is perpendicular to the plane of the figure. The similar calculation of spectrum for the sample with $N=1005$ and open boundary conditions is shown in \fref{fig:4} b). Chiral edge states of Majoraba fermions and band states of itinerant fermions form the spectrum of the system. In quasi irrational limit at $N=177$, when $q>N$, the spectrum contains predominantly chiral edge modes \fref{fig:4} c).

\subsection{Chiral Majorana fermion liquid, irrational flux $\phi=\frac{1}{\sqrt {8}}$}
\begin{figure}[tp]
    \centering{\leavevmode}
     \centering{\leavevmode}
   \begin{minipage}[h]{.475\linewidth}
\center{\includegraphics[width=\linewidth]{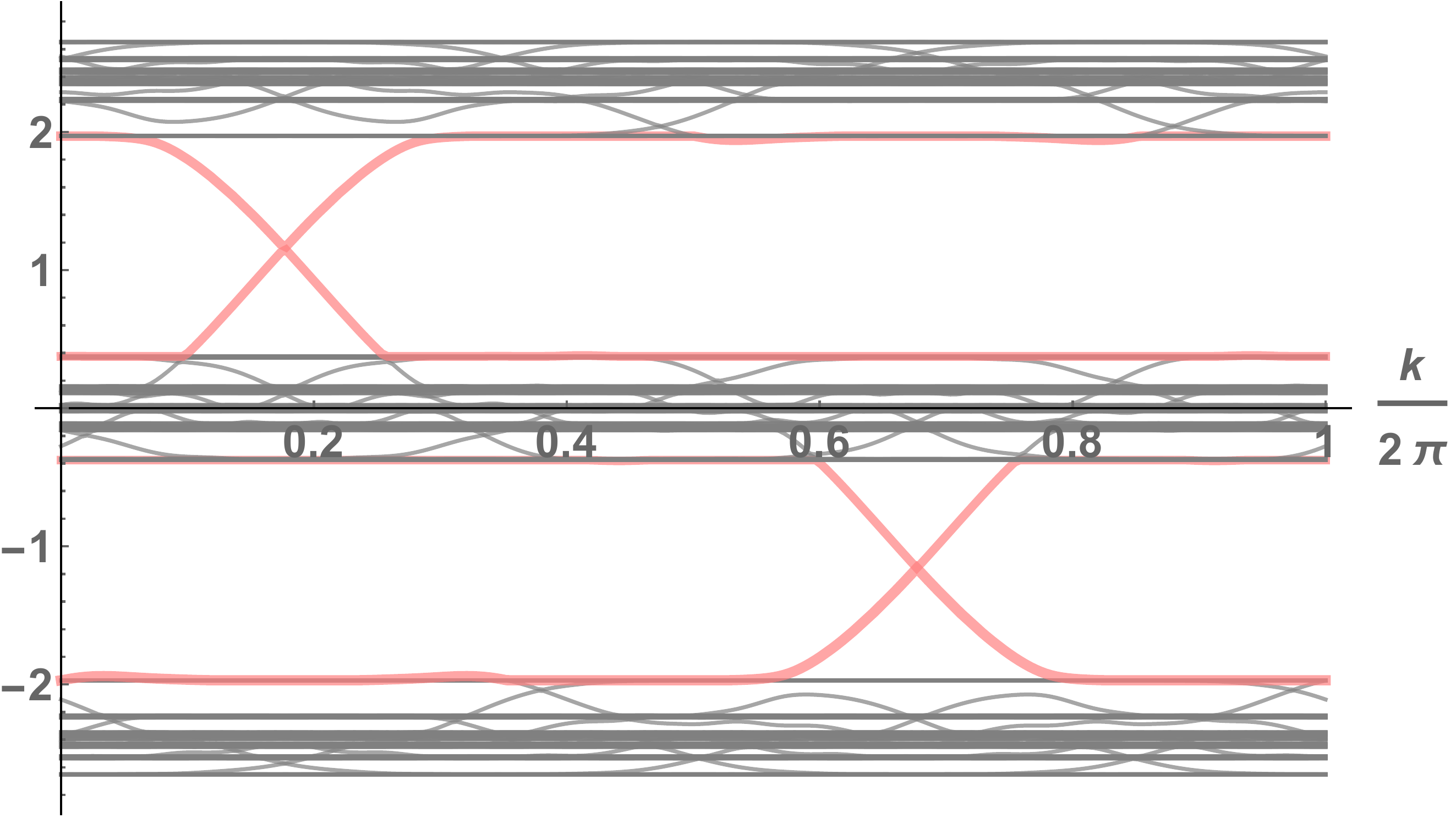}\\ a)}
\end{minipage}
\begin{minipage}[h]{.475\linewidth}
\center{\includegraphics[width=\linewidth]{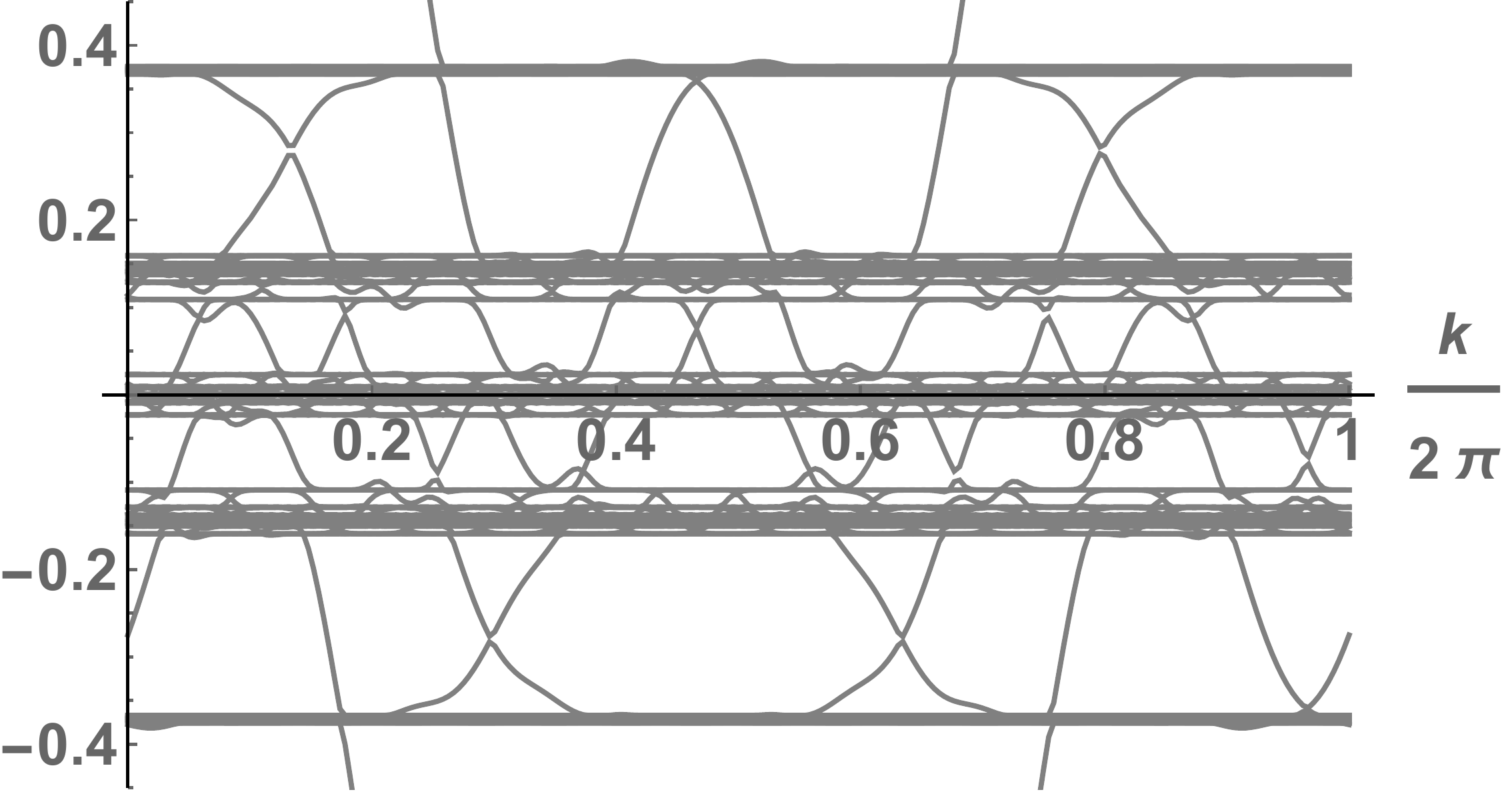}\\ b)}
\end{minipage}
\begin{minipage}[h]{.475\linewidth}
\center{\includegraphics[width=\linewidth]{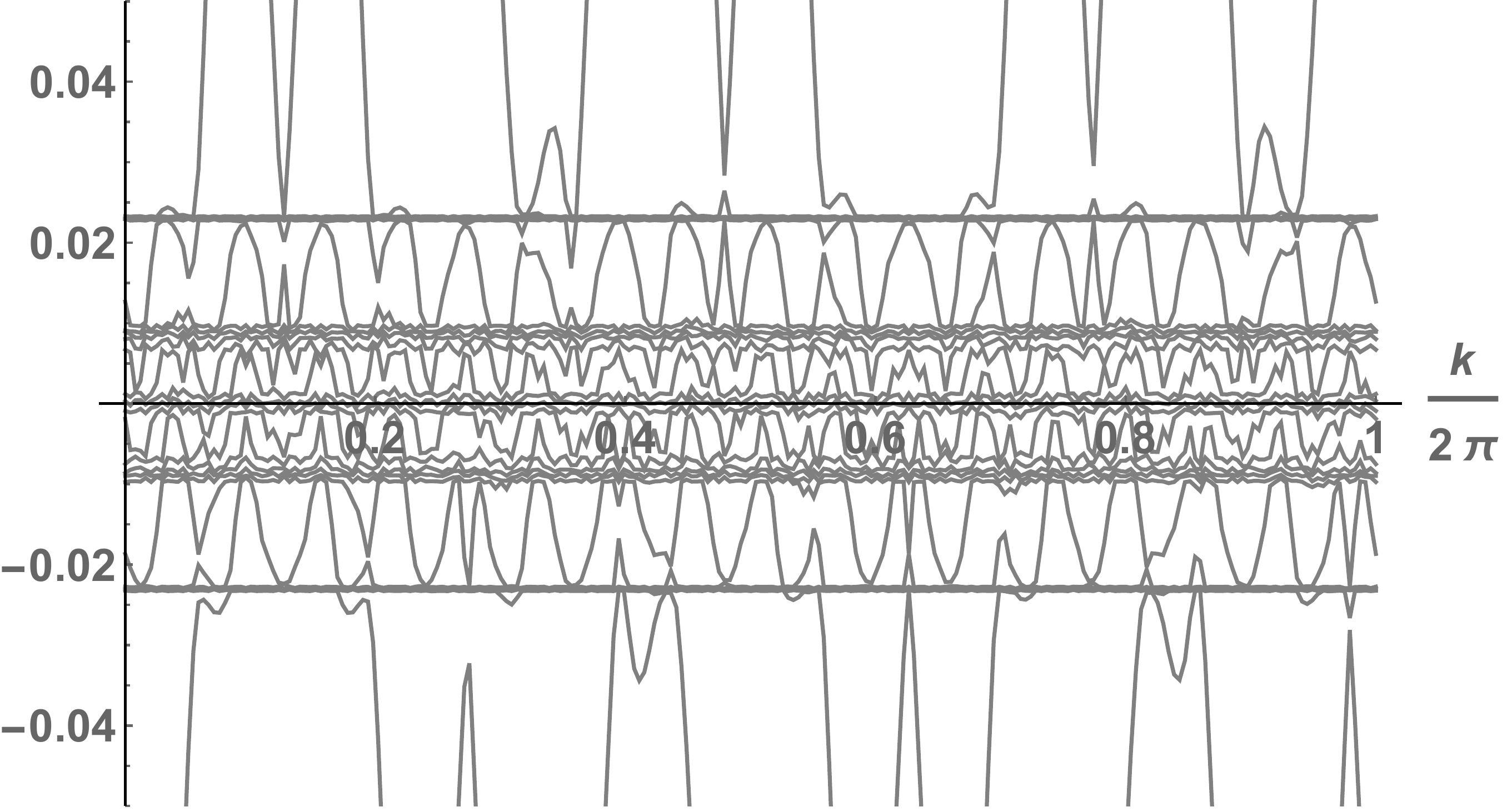}\\ c) }
\end{minipage}
\begin{minipage}[h]{.475\linewidth}
\center{\includegraphics[width=\linewidth]{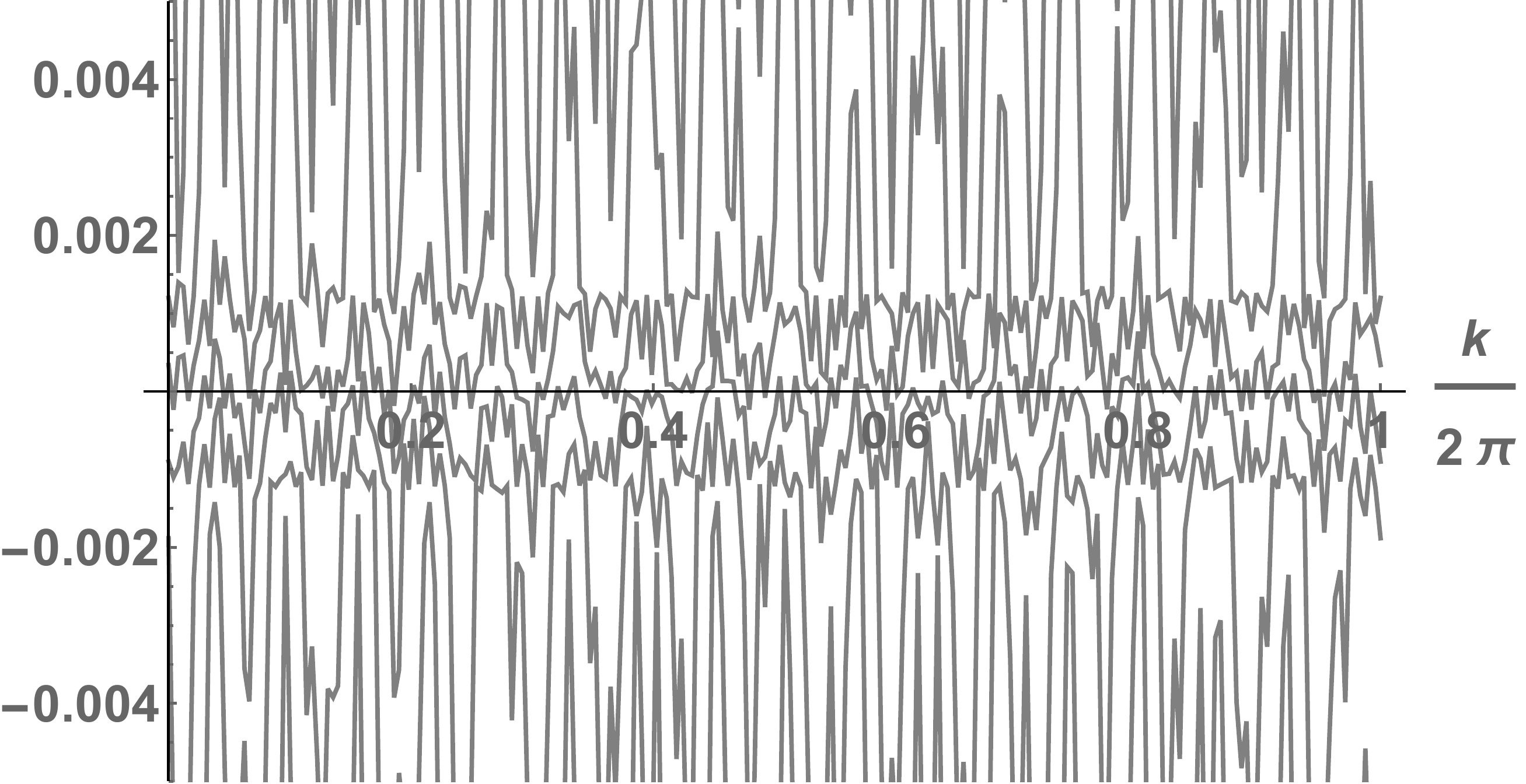}\\ c) }
\end{minipage}
\caption{(Color online)
Energy levels calculated for 
the flux $\phi=\frac{169}{478}$ at $t=1$ and $N=450$: a total spectrum a), a fine structure of the middle subband of the spectrum b), a hyperfine structure of a fine structure of middle subband c) and d).
  }
\label{fig:5}
\end{figure}

Numerical calculations of the excitation spectrum for a rational flux in Hofstadter strips with open boundary conditions were obtained for samples of sizes $N> q$, more precisely $\frac{N}{ q}$ is integer. When calculating the spectrum for a irrational flux which is approximated by rational numbers $\phi\simeq \frac{p}{q}$, we assume that $N < q$.

Consider the excitation spectrum, calculated at $\phi = \frac {1}{\sqrt 8}$ and $t=1$, analyzing the results of calculations presented in \fref{fig:5}. Three bands separated by wide gaps, connected by the chiral gapless edge modes, are topological ones with the Chern numbers $\{ 1,-2,1\}$, as in the case $q=3$ (see  in \fref{fig:1}a)). The number of gapless edge modes, indicated in red lines, determines the values of the Chern numbers of the isolated subbands. The number of these gapless edge modes is conserved which is confirmed by numerical calculations of the spectrum for a set of rational fluxes that correspond to an irrational flux. Each subband has a fine structure, a middle subband is shown in \fref{fig:5}b),c),d). The state of itinerant fermions are connected by the chiral edge modes. The density of the edge states increases when the energy approaches zero, which corresponds to half filling.
A hyperfine structure of the middle subband  with different scaling is shown in  \fref{fig:5} c).d). At the half filling  two component fermion liquid contains both itinerant and chiral Majorana fermion states. An anomalously large Hall conductivity is due to chiral Majorana localized states.

\section{Conclusions }

We studied the behavior of 2D spinless fermions in the framework of Hofstadter's model, focusing on the half filling.
A topological structure of the Hofstadter model is calculated for the CI state with a rational magnetic flux through the unit cell.
In the limit of weak flux per plaquette we obtained abnormal behavior of a Hall conductivity, when  the magnetic field is weak,  $\sigma_{xy}\simeq \frac{e}{2 H}$, that is a result of the large Chern numbers  $|C|\simeq 1/\phi$  of topological subband at zero energy. Such behavior of the system is determined by  two different liquid states of spinless fermions at half filling, namely, the liquid of itinerant fermions and chiral Majorane fermion liquid.
The chiral Majorana fermion liquid determines the anomalous behavior of the Hall conductivity in a weak magnetic field.

\end{document}